\documentclass[a4paper,11pt]{article}
\pdfoutput=1 
\usepackage{caption}
\usepackage{jcappub} 
                     
\usepackage{aasmacros}              
\usepackage{comment}
\usepackage{lscape}
\usepackage{verbatim}
\usepackage{comment}
\usepackage{array}
\usepackage{subcaption}
\usepackage{multirow}
\usepackage{graphicx}
\graphicspath{{graphics/}}
\usepackage{amsmath}
\usepackage{amssymb}
\usepackage{rotating} 
\usepackage{xspace}
\usepackage[dvipsnames]{xcolor}
\usepackage{booktabs}
\usepackage[normalem]{ulem}
\usepackage{adjustbox}
\usepackage{lipsum}  
\newcommand{\Msun}{{\rm M}_\odot}

\newcommand{\be}{\begin{equation}}
\newcommand{\ee}{\end{equation}}

\def\NB#1{{\color{blue}#1}}

\title{Dark matter annihilation signals from the Large Magellanic Cloud and its impact on the Milky Way}

\author[a]{Evan Vienneau,}
\author[b]{Evan Batteas,}
\author[b]{Addy J. Evans,}
\author[b,a]{Odelia V. Hartl,}
\author[a, d]{Nassim Bozorgnia,}
\author[b]{Louis E. Strigari}

\affiliation[a]{Department of Physics, University of Alberta,
CCIS 4-181, Edmonton, Alberta T6G 2E1, Canada}
\affiliation[b]{Department of Physics and Astronomy,
Mitchell Institute for Fundamental Physics and Astronomy,
Texas A$\&$M University, College Station, TX 77843, USA}
\affiliation[d]{Theoretical Physics Institute, University of Alberta,
CCIS 4-181, Edmonton, Alberta T6G 2E1, Canada}

\abstract{
We study the dark matter (DM) annihilation signals from the Large Magellanic Cloud (LMC) and the impact of the LMC on the DM annihilation signals from the Milky Way (MW) halo, using a MW-LMC analogue from the Auriga magneto-hydrodynamical simulations. We find that the gamma-ray signals from DM annihilation from the LMC rises above the MW foreground by a factor of greater than 100 for the s-wave velocity-independent annihilation model, as well as for the Sommerfeld, p-wave, and d-wave velocity-dependent models.  We derive upper limits on the annihilation cross section of DM particles in the LMC using Fermi-LAT data for all velocity-dependent cross section models. Bounds for d-wave annihilation are more stringent by $\sim 4-6$ orders of magnitude relative to previous bounds from dwarf galaxies, and for p-wave emission our bounds are $\sim 2-3$ orders of magnitude more stringent. We also demonstrate that the impact of the LMC on the DM annihilation signals from the MW halo is greatest for the p-wave and d-wave models towards the outer MW halo, while the impact is minimal for Sommerfeld and s-wave models. The LMC boosts the DM density and velocity distribution in the outer MW halo, both by bringing in high-speed DM particles and by accelerating the DM particles of the MW, affecting the DM annihilation signals from the MW for the p-wave and d-wave models. 
}

\begin{document}
\maketitle
\flushbottom

\section{Introduction}
\label{Introduction}

The Large Magellanic Cloud (LMC) is the largest and most massive satellite of the Milky Way (MW). It has just passed its first pericenter approach and is currently at a galactocentric distance of $\sim 50$~kpc, moving at a speed of $321 \pm 24$~km/s  
with respect to the MW~\cite{Besla:2007kf, Kallivayalil:2013xb, Patel:2017, Pietrzynski:2019}. Many properties of the LMC have been studied in detail using observational data. These include its diverse star formation history~\cite{Harris:2009}, the distribution and kinematics of its stellar populations~\cite{Werchan:2011sc,Wan:2020}, and the bar dominated dynamics of its inner disk~\cite{Jim:2023}. 

An important aspect regarding the LMC that has attracted significant interest in recent years is its impact on the MW, not only by bringing in dark matter (DM), gas, dust, and stars, but also by perturbing the MW's gravitational potential and reshaping its components. Studies based on observations have found evidence for an LMC-induced gravitational wake in the distribution of stars~\cite{Fushimi:2024, Conroy:2021, Cunningham:2020nlo}, a warping of the MW disk~\cite{Laporte:2016vuu} and a net reflex motion of the disk caused by the LMC~\cite{Petersen_2020, Peterson:2020, Gomez:2014tda}. These findings suggest a corresponding impact of the LMC on the DM component of the MW. Studies utilizing N-body simulations have shown that the LMC can induce asymmetric perturbations and large-scale over-densities in the MW's DM halo~\cite{GaravitoCamargo:2021tcp, Garavito-Camargo:2020lqm}.


The LMC and its impact on the MW halo provides a unique avenue for DM direct and indirect detection searches. Signals in direct detection experiments are sensitive to the local DM density and velocity distribution. The impact of the LMC on the DM distribution in the Solar region has been studied using cosmological simulations~\cite{Smith-Orlik:2023kyl, Reynoso-Cordova:2024xqz}, as well as idealized simulations of the MW-LMC system~\cite{Besla:2019xbx, Donaldson:2022}. These studies find that the LMC significantly boosts the local DM velocity distribution, strengthening direct detection exclusion limits by several orders of magnitude for low mass DM~\cite{Smith-Orlik:2023kyl, Reynoso-Cordova:2024xqz}. The boost in the high speed tail of the local DM velocity distribution arises from both DM particles originating from the LMC and MW particles that have been accelerated by the LMC’s gravitational influence~\cite{Besla:2019xbx, Smith-Orlik:2023kyl}.

The DM annihilation signals searched for in indirect detection experiments are sensitive to the DM distribution along the line of sight to the target. The astrophysical dependence of DM annihilation signals is encapsulated in the so-called $\mathcal{J}$-factor, which for velocity-independent annihilation models is an integral over the line of sight of the DM density squared. There is also the possibility that the annihilation cross section is velocity dependent, in which case the $\mathcal{J}$-factor also incorporates the pair-wise relative velocity distribution of DM along the line of sight. The most widely studied velocity-dependent models are p-wave~\cite{Kumar:2013iva}, d-wave~\cite{Giacchino:2013bta}, and Sommerfeld~\cite{Feng:2010zp} models, where the annihilation cross section, $\sigma_A v_{\rm rel}$, scales with  the DM relative velocity as $v_{\rm rel}^2$, $v_{\rm rel}^4$, and $v_{\rm rel}^{-1}$, respectively.

The LMC is an interesting indirect detection target since observational studies suggest that it harbors a significant DM halo, with recent estimates placing the total mass in the range of $\sim [1 - 3.5] \times 10^{11}$~M$_{\odot}$~\cite{Penarrubia:2016,Erkal:2019jji,Vasiliev:2021,Shipp:2021,Koposov:2023, Watkins:2024}. There have been no excesses detected from the LMC, but competitive bounds have been placed on the DM annihilation cross section using gamma-ray, radio and neutrino observations of the LMC \cite{Buckley:2015doa, Liang:2016bxu, Avrorin:2016yhw, Chan:2022wyj, Chen:2024zcl}. The LMC can also potentially impact the DM annihilation signals from the MW halo~\cite{Ackermann:2012, Chang:2018bpt, Calore:2021jvg, Eckner:2022swf}.
In particular, the injection of DM particles from the LMC into the MW halo, the acceleration of DM particles of the MW as a result of the passage of the LMC, and the LMC-induced asymmetric density perturbations could impact the expected DM annihilation or decay signals from the MW halo. Ref.~\cite{Eckner:2022swf} quantified this impact using idealized simulations and found that the dynamical response of the MW halo caused by the passage of the LMC can alter DM constraints at a level comparable to existing observational uncertainties in the DM density profile and total mass of the MW halo.

In this paper, we utilize the Auriga cosmological magneto-hydrodynamical simulations~\cite{Grand:2016mgo} to study the DM annihilation signals from the LMC, as well as the impact of the LMC on DM annihilation signals from the MW halo. We consider both velocity-independent and velocity-dependent DM annihilation models. The paper is structured as follows. In section~\ref{sec: simulations} we discuss the details of the simulations and the MW-LMC analogue that we use. In section~\ref{sec: distributions} we present the DM density profile and velocity distribution of the simulated LMC system and the MW analogue under the impact of the LMC. In section~\ref{sec: J-factors} we discuss the $\mathcal{J}$-factors of the LMC and the Milky Way analogues. In section~\ref{sec: fermi} we present an analysis of the Fermi-LAT data for the LMC. We present a brief discussion and our conclusions in section~\ref{sec: discussion}. In appendix~\ref{app: var}, we provide additional information on the variations in the MW $\mathcal{J}$-factor.

\section{Simulations}
\label{sec: simulations}

In this study, we use the magneto-hydrodynamical simulations of MW-mass halos from the Auriga project~\cite{Grand:2016mgo, Grand:2024xnm}. The Auriga simulation suite comprises 30 high-resolution cosmological zoom-in simulations of isolated MW-mass halos, selected from a 100$^3$ Mpc$^3$ periodic volume (L100N1504) originally part of the EAGLE project \cite{Schaye:2014tpa, Crain:2015}. These simulations were carried out using the moving-mesh code Arepo and implement a comprehensive galaxy formation subgrid model. This model incorporates metal cooling, black hole formation, AGN and supernova feedback, star formation, and background UV/X-ray photoionisation radiation \cite{Grand:2016mgo}. The simulations adopt cosmological parameters from Planck-2015 \cite{Planck:2015fie}: $\Omega_m = 0.307$, $\Omega_b = 0.048$, and $H_0 = 67.77$~km\,s$^{-1}$~Mpc$^{-1}$. We employ the standard resolution level (Level 4), which features a DM mass, $m_{\text{DM}} \sim 3 \times 10^5$~M$_{\odot}$, baryonic mass element, $m_b$ $\sim$ 5 $\times$ 10$^4$ M$_{\odot}$, and a Plummer equivalent gravitational softening length, $\epsilon= 370$~pc~\cite{Power:2002sw, Jenkins:2013raa}. The Auriga simulations reproduce the observed stellar masses, sizes, rotation curves, star formation rates, and metallicities of present day MW-mass galaxies.

\subsection{Selection of MW-LMC analogue}

To study the DM annihilation signal from the LMC and the MW halo itself, we use a simulated MW-LMC analogue that has properties similar to the observed MW-LMC system. The LMC's first pericenter approach occurred $\sim$ 50 Myr ago \cite{Besla:2007kf}. Accordingly, we will use the LMC's properties at or near its first pericenter passage. 
The present day stellar mass of the LMC from observations is $\sim 2.7 \times 10^9$~M$_\odot$~\cite{vanderMarel:2002kq}. 

We use a specific MW-LMC analogue selected from a collection of 15 MW-LMC analogues identified in ref.~\cite{Smith-Orlik:2023kyl}. These analogues were identified by requiring that their stellar masses and their distances from host at first pericenter approach match observations. 
The MW-LMC analogue that we study in this paper is the re-simulated halo 13 in ref.~\cite{Smith-Orlik:2023kyl}, corresponding to the Auriga 25 halo and its LMC analogue. This system has an LMC halo mass at infall of $3.2 \times 10^{11}$~M$_{\odot}$ and the virial mass of the MW analogue is $1.2 \times 10^{12}~\Msun$. The simulation was rerun with finer snapshots close to the LMC’s pericenter approach such that the average time between snapshots near pericenter in the new run is $\sim$ 10 Myr. More details regarding this system and its selection criteria are provided in ref.~\cite{Smith-Orlik:2023kyl}. 

To study the impact of the LMC on the MW, we consider two snapshots for the MW-LMC system: isolated MW (\emph{Iso.}) and present day MW-LMC (\emph{Pres.}). Iso.~takes place when the MW and LMC analogues have the largest separation, at the first apocenter of the LMC before infall. Pres.~is the closest snapshot to the present day separation of the observed MW and LMC system. At the Pres.~snapshot, the LMC analogue has a velocity of 317~km/s and a distance of $\sim 50$~kpc from the host MW analogue, closely matching the observed values.

\subsection{Choosing the Solar position}
\label{sec: Solar}

In the simulations, the Solar position is not determined a priori. We choose the Solar position such that the position of the LMC analogue matches its correct on-sky location ($l =79.5^\circ$, $b =-32.9^\circ$), with the galactic center at $l =0^\circ$, $b =0^\circ$. 

We define a reference frame with the origin at the galactic center and the $z$-axis perpendicular to the stellar disk. The orientation of the stellar disk is identified by determining the average angular momentum vector of all stars within 10~kpc of the galactic center, specifying the $z$-axis. We then sample over $\sim$ 2$\times$10$^5$ Solar positions contained within a torus which is aligned with the stellar disk and has a radius that extends between 7 to 9~kpc from the galactic center, and a height extending from $-1$ to 1~kpc below and above the stellar disk, respectively. We choose the Solar position for which the resulting on-sky position of the LMC analogue with respect to the MW center best matches the observed position. We find the LMC analogue position to be within 1 degree of the target latitude and longitude. The resulting distance between the best Solar position and the LMC analogue is $\sim 51$~kpc.

\section{Dark matter distribution}
\label{sec: distributions}

In this section we present the DM density and pair-wise relative speed distribution of the LMC analogue and the MW analogue, extracted from the simulations.

\begin{figure}[t]
    \centering
    \includegraphics[width=1\linewidth]{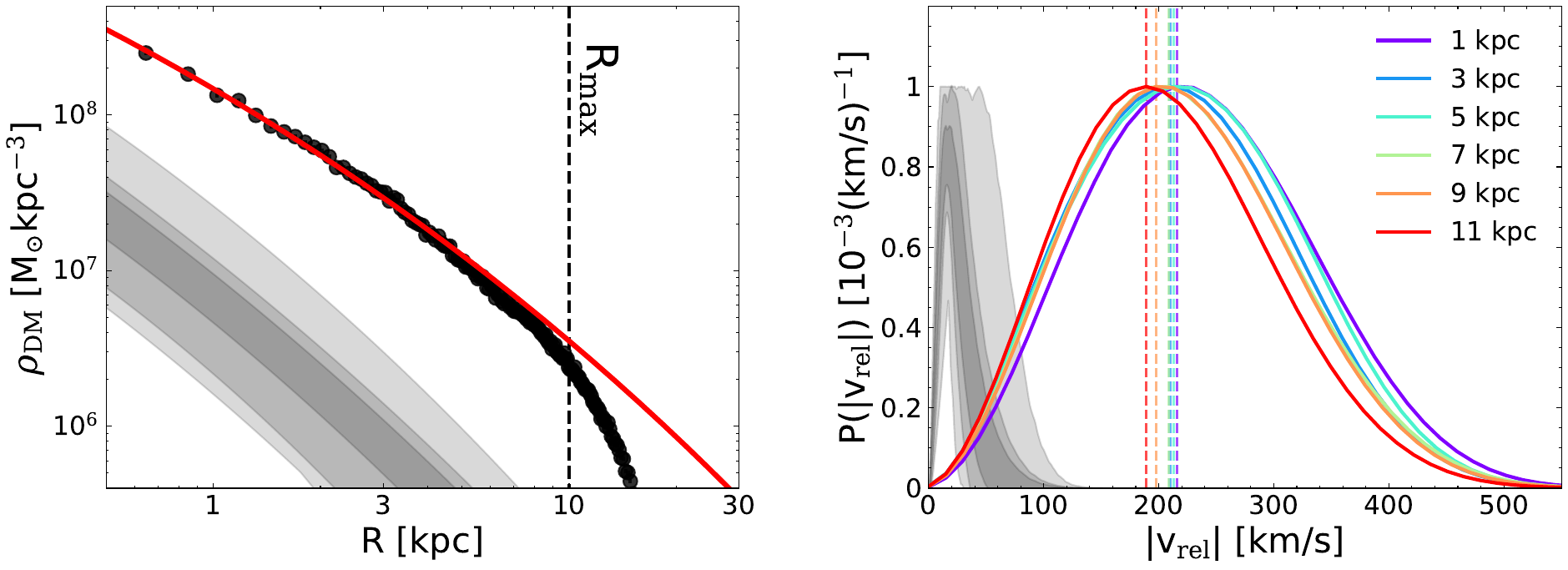}
    \caption{Left: DM density profile of the LMC analogue in the present day snapshot (black points) and its best fit Einasto profile (red curve). The vertical dashed line indicates the radius of LMC's maximum circular velocity, $R_{\rm max}$. The 1$\sigma$ Poisson error bars in the density profile are smaller than the  size of the data points. 
    Right: The DM relative speed distribution of the LMC analogue at various radial distances from the LMC's center. Vertical lines are placed at the peak of each distribution. The three shaded  gray bands in the left and right panels contain 100$\%$ (light gray), 68$\%$ (darker gray), and 38$\%$ (darkest gray) of the DM density profiles and relative speed distributions around the mean, respectively, of the other subhalos in the MW which have more than 500 DM particles.
    }\label{density_veldist}
\end{figure}

The left panel of Figure~\ref{density_veldist} shows the DM density profile of the LMC analogue at the present day snapshot (black data points). The density profile is calculated using concentric radial shells with non-uniform width, each containing 1000 DM particles. For comparison, the shaded gray bands show the DM density profiles of all other MW subhalos, which have more than 500 DM particles. We fit the DM density profile of the LMC with an Einasto density profile given by
\begin{equation} 
\rho(r)=\rho_{-2}\exp \left( -\frac{2}{\alpha}\left[ \left( \frac{r}{r_{-2}}\right)^\alpha-1\right]\right),
\label{NFW}
\end{equation}
where $r_{-2}$ and $\rho_{-2}$ represent the radius and density at which the logarithmic  slope of the density profile is equal to $-2$, and $\alpha$ is a parameter specifying the curvature of the density profile. 
We fit the Einasto density profile to the simulation data up to the radius of LMC's maximum circular velocity, $R_{\rm max}$. As it can be seen from the left panel of figure~\ref{density_veldist}, the DM density profile of the LMC decreases faster than an Einasto density profile at radii larger than $R_{\rm max}$, due to tidal stripping. The red curve shows the best fit Einasto profile, with best fit parameters, $r_{-2}=5.65$ kpc, $\rho_{-2}=9.36\times10^{6}$ M$_{\odot}$, and $\alpha=0.277$. We note that previous work \cite{Gao:2007gh} has found that relaxed cold DM halos at redshift $z \simeq0$ are well-fitted by an Einasto profile with $\alpha \sim 0.16$. However, we find that this $\alpha$ value does not yield a fit that accurately captures the density profile of the LMC compared to the fit we obtain when leaving $\alpha$ as a free parameter. This may be due to the fact that the LMC is on its first orbit around the MW and its halo is therefore most likely not relaxed.

The right panel of figure~\ref{density_veldist} shows the DM pair-wise relative speed distribution of the LMC at various radial distances from the LMC's center. The relative speed distribution can be computed from the relative velocity distribution, $P_{\bf x}({{\bf v}_{\rm rel}})$, at position ${\bf x}$ in a halo by,
\begin{equation}
P_{\bf x}({|{\bf v}_{\rm rel}}|)=v_{\rm rel}^2 \int P_{\bf x}({{\bf v}_{\rm rel}})\, d\Omega_{{\bf v}_{\rm rel}},
\end{equation}
where $d\Omega_{{\bf v}_{\rm rel}}$ is an infinitesimal solid angle along the direction ${\bf v}_{\rm rel}$. 

The relative speed distributions in the right panel of figure~\ref{density_veldist} are computed in spherical shells of 2~kpc width centered at a radial distance ranging from 1~kpc to 11~kpc from the LMC's center. As expected, the peak of the distribution shifts to lower speeds as the distance from the center of the LMC increases. The shaded gray bands encapsulate the DM relative speed distributions of all other MW subhalos, which have more than 500 DM particles. The distributions are normalized to peak at 1.

It is evident from figure~\ref{density_veldist} that the LMC is unique compared to other MW subhalos, due to its considerably larger mass. As a result, its DM density is larger and its DM relative speed distribution peaks at a higher speed, compared to other MW subhalos.

\begin{figure}[t]
    \centering
    \includegraphics[width=0.6\linewidth]{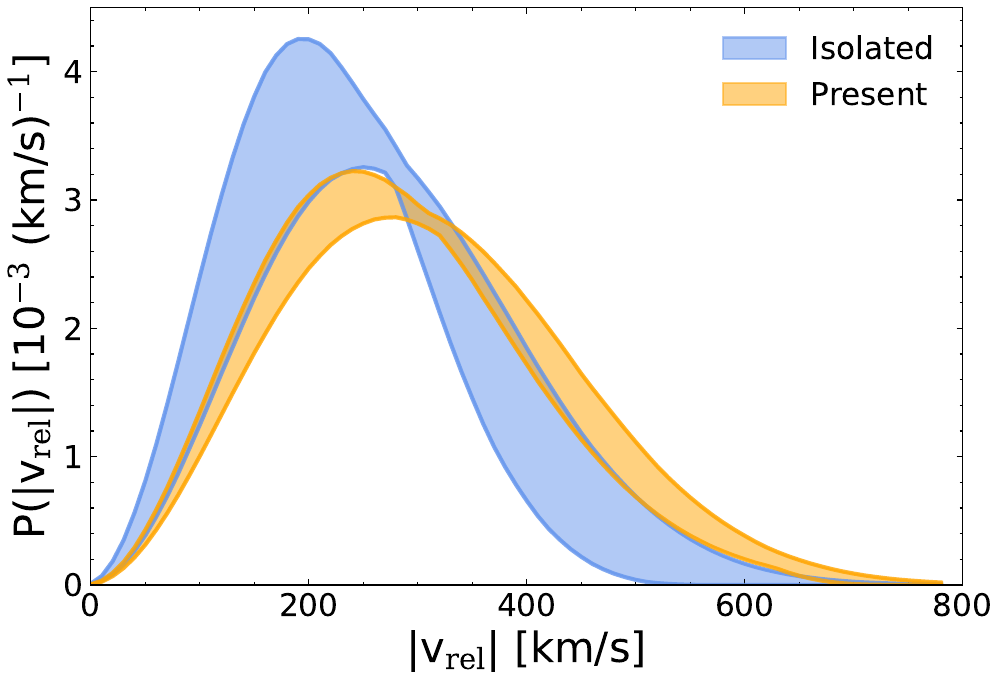}
    \caption{DM relative speed distributions of the MW analogue computed in 1 kpc thick radial shells from 2 - 50 kpc. The blue and orange shaded regions encapsulate all of these curves for the $\textit{Iso.}$ snapshot and $\textit{Pres.}$ snapshot (including DM particles from both the MW and LMC), respectively.}\label{relvel_iso_pres}
    \label{fig:veldist}
\end{figure}

Next, we explore how the LMC impacts the DM pair-wise relative speed distribution of the MW analogue. Figure~\ref{relvel_iso_pres} shows a comparison of the DM relative speed distribution of the isolated MW before the infall of the LMC ({\it Iso.}~snapshot) and that of the present day MW-LMC ({\it Pres.}~snapshot) including the DM particles from both the MW and LMC. The relative speed distributions are calculated in 1~kpc thick spherical shells centered on the MW center, with radii in the range of $[2 - 50]$~kpc. The blue and orange shaded bands encapsulate all these curves for the $\textit{Iso.}$ and $\textit{Pres.}$ snapshots, respectively. The distributions are normalized such that $\int dv_{\rm rel}\, P_{\bf x}({|{\bf v}_{\rm rel}}|)=1$. It is clear that the LMC shifts the DM relative speed distribution in the halo to higher speeds. We also find that the DM relative speed distribution is shifted more so at larger galactocentric distances; this is expected since the LMC is at a galactocentric distance of $\sim 50$~kpc.

\section{$\mathcal{J}$-factors}
\label{sec: J-factors}

In this section we discuss the $\mathcal{J}$-factors of the LMC analogue as well as the MW analogue under the influence of the LMC. The $\mathcal{J}$-factor encapsulates the astrophysical dependence of the DM annihilation signal. For velocity-dependent annihilation models, the $\mathcal{J}$-factor depends on both the DM density profile and the DM pair-wise, or relative velocity distribution. We follow closely the formalism
in refs.~\cite{Board:2021bwj,Blanchette:2022hir, Vienneau:2024xie} to compute the $\mathcal{J}$-factors for different DM annihilation models. 
The DM annihilation cross section, $\sigma_A$, averaged over the DM relative velocity distribution, $P_\textbf{x}(\textbf{v}_{\rm rel})$, at position $\textbf{x}$ in a halo is given by
\begin{equation}
\langle \sigma_A v_{\rm rel} \rangle (\textbf{x}) = \int d^3 \textbf{v}_{\rm rel} P_\textbf{x}(\textbf{v}_{\rm rel})(\sigma_A v_{\rm rel})\, ,
\label{eq:annihilation-cross-section}
\end{equation}
where ${\bf v}_{\rm rel}={\bf v}_1 - {\bf v}_2$ is the DM relative velocity, with $v_{\rm rel}\equiv|{\bf v}_{\rm rel}|$.

In general, $\sigma_Av_{\text{rel}}$ can be parametrized as $\sigma_Av_{\text{rel}}$ = ($\sigma_Av_{\text{rel}}$)$_0$($v_{\text{rel}}$/c)$^n$, and depends on the relative DM velocity. ($\sigma_Av_{\text{rel}}$)$_0$ is the velocity-independent component of the annihilation cross section, and $n$ depends on the specific DM annihilation model. We consider the following models: $n=0$ (s-wave annihilation), $n=2$ (p-wave annihilation), $n = 4$ (d-wave annihilation), and $n=-1$ (Sommerfeld-enhanced annihilation). The $\mathcal{J}$-factor is defined as~\cite{Board:2021bwj, Boddy:2019wfg},

\begin{equation} 
{\mathcal J} (\theta) = \int d \ell \, \frac{\langle \sigma_A v_{\rm rel} \rangle}{(\sigma_A v_{\rm rel})_0}  \left[\rho (r(\ell, \theta))\right]^2 = \int d \ell \int d^3 {\bf v}_{\rm rel} P_{{\bf x}} ({\bf v}_{\rm rel}) ~\left(\frac{{v}_{\rm rel}}{c}\right)^n~ \left[\rho (r(\ell, \theta))\right]^2\, ,
\label{eq:Jfactor}
\end{equation} 
where $\ell$ is the line of sight distance from the Sun to a point in the target, $\theta$ is the opening angle between the line of sight $\ell$ and the distance $D$ from the Sun to the target of interest, and $r^2(\ell, \theta) = \ell^2 + D^2 - 2\ell D \cos{\theta}$ is the square of the radial distance measured from the center of the target. Different DM annihilation models correspond to different velocity moments of the relative velocity distribution~\cite{Board:2021bwj},
\begin{equation} 
\mu_n(\textbf{x}) \equiv\int d^3 {\textbf{v}}_{\rm rel} P_{{\bf x}} ({\bf v}_{\rm rel})\,{v}_{\rm rel}^n \,,
\end{equation} 
where $\mu_n(\textbf{x})$ is the $n$-th moment of the relative velocity distribution, $P_{{\bf x}} ({\bf v}_{\rm rel})$. In terms of the velocity moments, the $\mathcal{J}$-factor can be written as
\begin{equation} 
{\mathcal J} (\theta) = \int d \ell \, \left[\rho \left(r\left(\ell, \theta\right)\right) \right]^2 \, \left(\frac{\mu_n\left( r\left(\ell, \theta\right) \right)}{c^n}\right)\,.
\label{eq:Jfactor}
\end{equation}

The expected gamma-ray flux from DM annihilation is, in general, proportional to ${\mathcal J}$ and can be written as,

\begin{equation}
\frac{d \Phi_\gamma}{dE} = \frac{\left(\sigma_A v_{\rm rel}\right)_0}{8 \pi m_{\rm DM}^2}\frac{dN_\gamma}{dE} \,\mathcal{J},
\label{eq:flux}
\end{equation}
where $m_{\rm DM}$ is the DM particle mass, and $dN_\gamma/dE$ is the gamma-ray energy spectrum produced per annihilation.

The $\mathcal{J}$-factors are computed separately for the smooth halo of the MW and for the LMC. For the MW smooth halo component, the local DM density around each DM particle is estimated using a Voronoi tessellation. Then the DM relative velocity distribution is computed at each point on a spherical mesh using the nearest 500 DM particles~\cite{Piccirillo:2022qet}. These values are then interpolated to obtain an estimate for the DM relative velocity distribution at every point in the smooth halo. For the LMC analogue, the local DM density is found using the best fit Einasto density profile within $R_{\rm max}$ and a Voronoi tessellation outside of $R_{\rm max}$. The DM density profile is computed such that each radial shell of non-uniform width has at least 1000 DM particles. The DM relative velocity distribution is also calculated in these consecutive spherical shells and the average is applied to all DM particles in the LMC analogue. The integrand of the $\mathcal{J}$-factor for each DM particle is formed by the product of the local DM density squared and the moment of the local relative velocity distribution scaled by $1/c^n$. The $\mathcal{J}$-factors are then computed by integrating the contribution from all DM particles along the given line of sight in uniform patches on the sky.

Note that in calculating the $\mathcal{J}$-factors, in addition to DM particles that are bound to the LMC, we also include the DM particles bound to the MW that are within the $R_{\rm max}$ of the LMC but are not bound to it. In a previous work~\cite{Vienneau:2024xie}, we showed that the presence of these fast moving MW DM particles in the region of the Sagittarius dwarf spheroidal galaxy can boost the p-wave and d-wave $\mathcal{J}$-factors by a factor of $\sim 2 - 30$ and $\sim 20 - 4000$, respectively. In a follow up study~\cite{Hartl:2025mtb}, we further quantified this effect by computing the boost for a large array of subhalos. We found that the boost is largest for small subhalos closest to the galactic center. In agreement with ref.~\cite{Hartl:2025mtb}, we find that the impact of unbound particles is negligible for the LMC analogue, due to its large size and distance from the MW.

In sections~\ref{sec: LMCJ} and \ref{sec: MWJ} we discuss two different effects, the $\mathcal{J}$-factor from the LMC itself and the impact of the LMC on the $\mathcal{J}$-factor from the MW halo.

\subsection{The LMC $\mathcal{J}$-factors}
\label{sec: LMCJ}

\begin{figure}[t]
    \centering
    \includegraphics[width=0.85\linewidth]{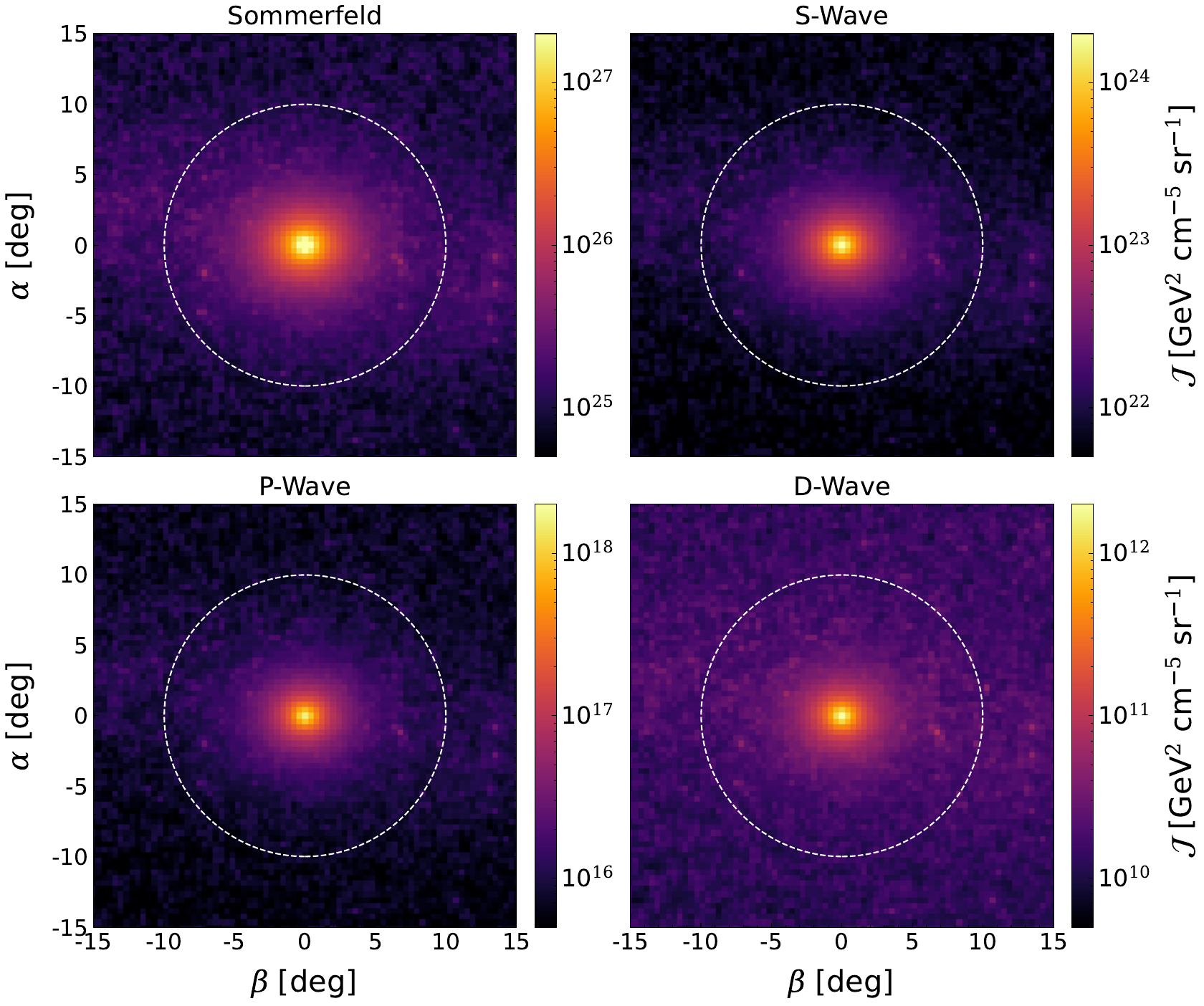}
    \caption{$\mathcal{J}$-factor sky-maps centered on the LMC analogue for the Sommerfeld (top left), s-wave (top right), p-wave (bottom left) and d-wave (bottom right) models. $\alpha$ and $\beta$ are the polar and azimuthal angles in a randomly defined spherical coordinate system centered on the LMC analogue, with the $z$-axis pointing from the position of the Sun to the LMC analogue. The dashed white circle specifies the angular extension of the $R_{\rm max}$ of the LMC.}\label{lmc_map}
\end{figure}

Figure~\ref{lmc_map} shows the $\mathcal{J}$-factor maps for a $30 \times 30$ degree region around the LMC analogue, for the different velocity-dependent DM annihilation models. For each panel we use a pixel size of 0.4 degrees on each side. As expected, the $\mathcal{J}$-factors are highest for the Sommerfeld-enhanced model and smallest for d-wave.

\begin{figure}[t]
    \centering
        \includegraphics[width=0.6\linewidth]{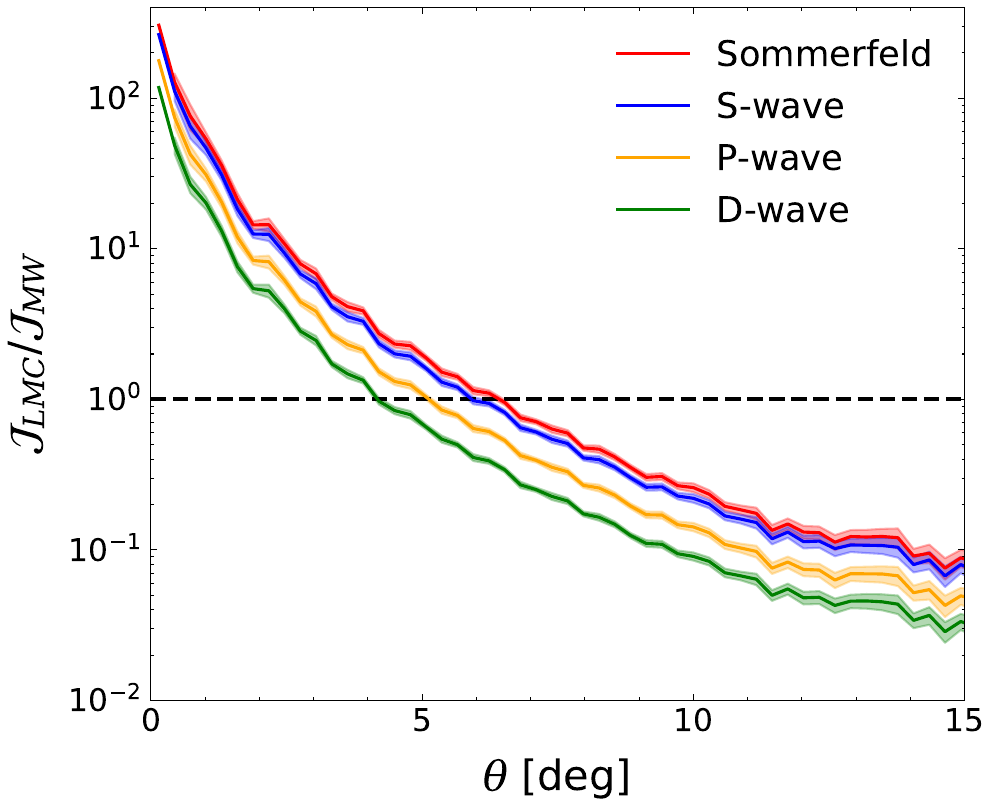}
    \caption{Ratio of the $\mathcal{J}$-factors for the LMC analogue and the smooth component of the host MW halo. The red, blue, orange, and green curves show the results for the Sommerfeld, s-wave, p-wave, and d-wave DM annihilation models, respectively. The shaded color bands indicate the standard error of the mean in each concentric circular angular bin on the sky. The dashed black line is placed at the point where the LMC emission equals the smooth MW emission.}\label{lmc_jfac}
\end{figure}

Figure \ref{lmc_jfac} shows the ratios of the $\mathcal{J}$-factor of the LMC analogue and the smooth component of the host MW halo along the line-of-sight to the LMC analogue. The red, blue, yellow and green curves show this ratio for the Sommerfeld, s-wave, p-wave, and d-wave models, respectively. $\mathcal{J}$-factors are computed in evenly spaced angular bins on the sky centered on the LMC analogue. Note that the subhalo contribution along the line of sight to the LMC analogue is negligible, and we have therefore not included it in the calculation of the $\mathcal{J}$-factor of the host MW halo in this figure. The emission from the LMC rises above the MW foreground by a factor of more than a 100 for all annihilation models. The angular distance at which the LMC emission falls below the MW emission is largest for the Sommerfeld model and then decreases slightly from s, to p to d-wave.

\subsection{The MW $\mathcal{J}$-factors}
\label{sec: MWJ}

\begin{figure}[t]
    \centering
    \includegraphics[width=0.95\linewidth]{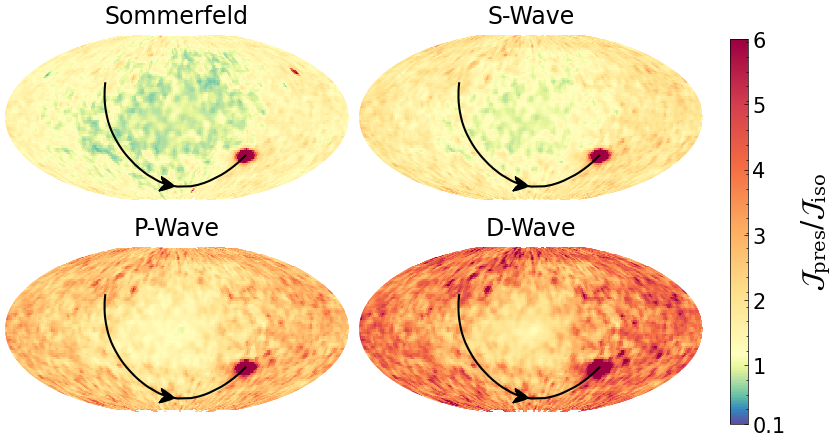}\hspace{5pt}
    \caption{All-sky maps showing the ratio of the MW halo $\mathcal{J}$-factor for the $\textit{Pres.}$ and $\textit{Iso.}$ snapshots, centered on the MW center and shown in a Mollweide projection for the Sommerfeld (top left), s-wave (top right), p-wave (bottom left) and d-wave (bottom right) models. The black arrowed curve in each panel indicates the orbit of the LMC analogue around the MW analogue. }\label{mw_present_isolated_jfac}
\end{figure}

To demonstrate the impact of the LMC on the MW halo, in figure~\ref{mw_present_isolated_jfac} we show all-sky maps of the ratio of the $\mathcal{J}$-factor of the MW halo for the $\textit{Pres.}$ (present day MW under the impact of the LMC) and $\textit{Iso.}$ (MW before LMC's infall) snapshots, for the different velocity-dependent DM annihilation models. For each panel, we use a pixel size of 2 degrees on each side. The maps are centered on the MW center and the orbit of the LMC from infall to the present day is shown as a black arrowed curve in each map. 

The ratio maps in figure~\ref{mw_present_isolated_jfac} include two effects due to the LMC: the DM particles originating from the LMC and the change in the distribution of MW DM particles due to the orbit of the LMC. The LMC causes overdensities in the DM particles in the outer MW halo. It also shifts the DM relative speed distribution of the MW towards higher speeds, as shown in figure~\ref{fig:veldist}. This boost is larger for particles in the outer halo compared to the inner halo. The interplay between the impact of the LMC on the DM density and the relative velocity distribution leads to the features seen in the maps in figure~\ref{mw_present_isolated_jfac}.
From the Sommerfeld map shown in the top left panel of figure~\ref{mw_present_isolated_jfac}, there is evidence for a  reduction of the $\mathcal{J}$-factor in the inner halo, while the $\mathcal{J}$-factor ratio is $\sim$ 1  in the outer halo. This is due to a competing impact of the DM density and relative velocities in the Sommerfeld case. In particular, the increase in the DM density in the outer halo boosts the $\mathcal{J}$-factor, and at the same time the high DM relative velocities reduce the $\mathcal{J}$-factor. 

The density boost is minimal in the inner halo, allowing for the $\mathcal{J}$-factor reduction due to the DM relative velocity to be noticeable. For the s-wave model shown in the top right panel of figure~\ref{mw_present_isolated_jfac}, the $\mathcal{J}$-factor ratio is $\sim$ 1 in the inner halo and slightly boosted in the outer halo, once again due to the larger DM overdensities in the outer halo.

The impact of the LMC becomes more noticeable for p-wave and d-wave as shown in the bottom panels of figure~\ref{mw_present_isolated_jfac}, where the $\mathcal{J}$-factor can be boosted by up to a factor of $\sim 4$ and $\sim 6$, respectively. The $\mathcal{J}$-factors for these annihilation models are strongly sensitive to the high speed tail of the DM relative velocity distribution, which is boosted due to the LMC. This is more significant in the outer halo where the increase in the DM density and relative velocity distribution is the largest. 

Comparing the four panels of figure~\ref{mw_present_isolated_jfac} suggests that the most prominent impact of the LMC on the MW $\mathcal{J}$-factors is in boosting the DM relative velocity distribution as opposed to significantly changing the DM density.

It is important to note that when computing the $\mathcal{J}$-factor in the direction of the outer halo, the immediate vicinity around the Solar position significantly impacts the magnitude of the resulting $\mathcal{J}$-factor. This is because for these specific lines of sight, the region around the Solar position has the highest DM density through which the lines of sight pass. This suggests that the resulting boost in the $\mathcal{J}$-factor of the $\textit{Pres.}$ snapshot with respect to the $\textit{Iso.}$ snapshot due to the LMC strongly depends on the specific DM density and relative velocity distribution around the chosen Solar position. We will discuss the variations in the MW $\mathcal{J}$-factor boost due to changing the Solar position in appendix~\ref{app: var}.

Figure~\ref{mw_present_nolmcparticles_jfac} shows all-sky maps of the ratio, $\mathcal{J_{{\rm MW}+{\rm LMC}}}/\mathcal{J_{\rm MW}}$, for the  MW halo in the $\textit{Pres.}$ snapshot, where $\mathcal{J_{{\rm MW}+{\rm LMC}}}$ includes the DM particles originating from both the MW and LMC, while $\mathcal{J_{\rm MW}}$  includes only the DM particles from the MW at the present day. This map shows how the DM particles originating from the LMC impact the $\mathcal{J}$-factor of the MW. There is minimal impact for s-wave and Sommerfeld models, whereas a tail of enhanced emission along the orbit of the LMC analogue is visible for the p-wave and d-wave models. This suggests that the density enhancement in the MW due to the DM particles originating from the LMC is minimal, while the boost in the relative DM velocity distribution is more substantial, supporting the conclusions drawn from figure~\ref{mw_present_isolated_jfac}. Moreover, comparing the magnitude of the $\mathcal{J}$-factor ratios in figures~\ref{mw_present_isolated_jfac} and \ref{mw_present_nolmcparticles_jfac} shows that the impact of the LMC gravitationally boosting the DM particles of the MW is more significant on the Milky Way $\mathcal{J}$-factor than the impact of the high speed DM particles originating from the LMC.

\begin{figure}[t]
    \centering
    \includegraphics[width=0.95\linewidth]{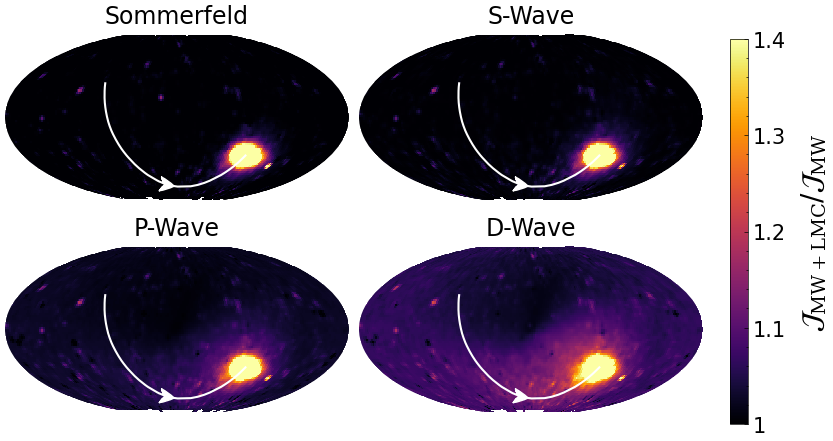}\hspace{5pt}
    \caption{All-sky maps showing the ratio of the MW halo $\mathcal{J}$-factor in the $\textit{Pres.}$ snapshot when all DM particles from the MW and the LMC are included ($\mathcal{J_{{\rm MW}+{\rm LMC}}}$) and when only the MW DM particles are included ($\mathcal{J_{\rm MW}}$). The maps are centered on the MW center and shown in a  Mollweide projection for the Sommerfeld (top left), s-wave (top right), p-wave (bottom left) and d-wave (bottom right) models. The white arrowed curve in each panel indicates the orbit of the LMC analogue around the MW analogue. }\label{mw_present_nolmcparticles_jfac}
\end{figure}

\section{Fermi analysis of the LMC}
\label{sec: fermi}

In this section we present an analysis of the Fermi-LAT data using our LMC analogue as a template. We include all the known astrophysical components of the LMC emission in addition to the DM source, and examine the impact of different assumptions for the DM mass and annihilation cross section. We compare the sensitivity to independent methods for constraining the velocity-dependent DM annihilation cross section.

Gamma-ray detection of the LMC was first reported by the Fermi-LAT Collaboration in 2010~\cite{Fermi2010}, after being marginally detected by EGRET several decades before~\cite{EGRET1992}. The extended emission coincident with the galaxy was first attributed only to the massive star-forming region 30 Doradus. With the accumulation of Fermi-LAT data, the emission was later separated into five distinct extended sources, with one region ``E0", later labeled as ``LMC-Galaxy" in subsequent catalogs, coinciding with the disk of the LMC itself~\cite{Fermi2016}. The emission from the disk was attributed to interactions of cosmic rays with the ISM of the galaxy.

\par In addition to the gamma-rays generated from cosmic-ray interactions, several authors have tested for the presence of a possible DM component contributing to the observed emission from the LMC. Using 5 years of data, a marginal, $< 2\sigma$ excess centered on the LMC was reported~\cite{Buckley2015}. This excess emission  was found in addition to the five previously known extended LMC sources. Using independent models to describe the astrophysical backgrounds of the system~\cite{Calore2022b}, combined with the LMC sources found in the 2FGL data release, a fit for the DM component was found for a given mass and cross section. However, these results were still formally statistically consistent with null detections~\cite{Eckner2023}. As pointed out in refs.~\cite{Buckley2015, Calore2022b}, there are significant degeneracies between the astrophysically attributed FGL source labeled ``LMC-Galaxy", centered on the galaxy with a radius of 3$^\circ$, and a possible DM signature, making the region difficult to reliably model when including a possible DM component.

\par Here we use the $\mathcal{J}$-factor maps generated from the simulations as a template to perform an updated, extended source binned likelihood analysis on the LMC region. To analyze the data we use Fermitools 2.2.0 with $\texttt{FermiPy}$. For our data selection, we use 16.57 years of data, corresponding to mission elapsed times between 239557417 s and 762177465 s. We select $\texttt{FRONT}$ and $\texttt{BACK}$ converting (evclass == 128 and evtype  == 3  ) events with energies between 500 MeV and 1 TeV, which is appropriate for the DM mass ranges that we are examining. We apply the suggested $\texttt{(DATA\_QUAL>0) \&\& (LAT\_CONFIG==1) }$filter to ensure quality data and a zenith cut of $z_{\text{max}} = 90^{\circ}$ to filter out gamma-ray contamination from the Earth's limb. We consider a $10^{\circ}  \times 10^{\circ}$ ROI centered on the LMC. This radius is large enough to encompass the majority of the DM emission from the simulated LMC analogue. We use the $\texttt{MINUIT}$ optimizer within $\texttt{gtlike}$ with a $0.01^{\circ}$ angular pixelation for our likelihood maximization. We account for all known sources in the $\texttt{4FGL}$ source catalog within the $10^{\circ}  \times 10^{\circ}$ region. For simplicity, and to best understand the impact of our DM source on the emission from the region, we fix all non-DM sources within the region to their best-fit 4FGL catalog values. For the interstellar emission model we use the recommended $\texttt{gll\_iem\_v07.fits}$, and for the isotropic emission we use $\texttt{iso\_P8R3\_SOURCE\_V3\_v1.txt}$. 

\begin{figure}[t]
    \centering
    \includegraphics[width=\linewidth]{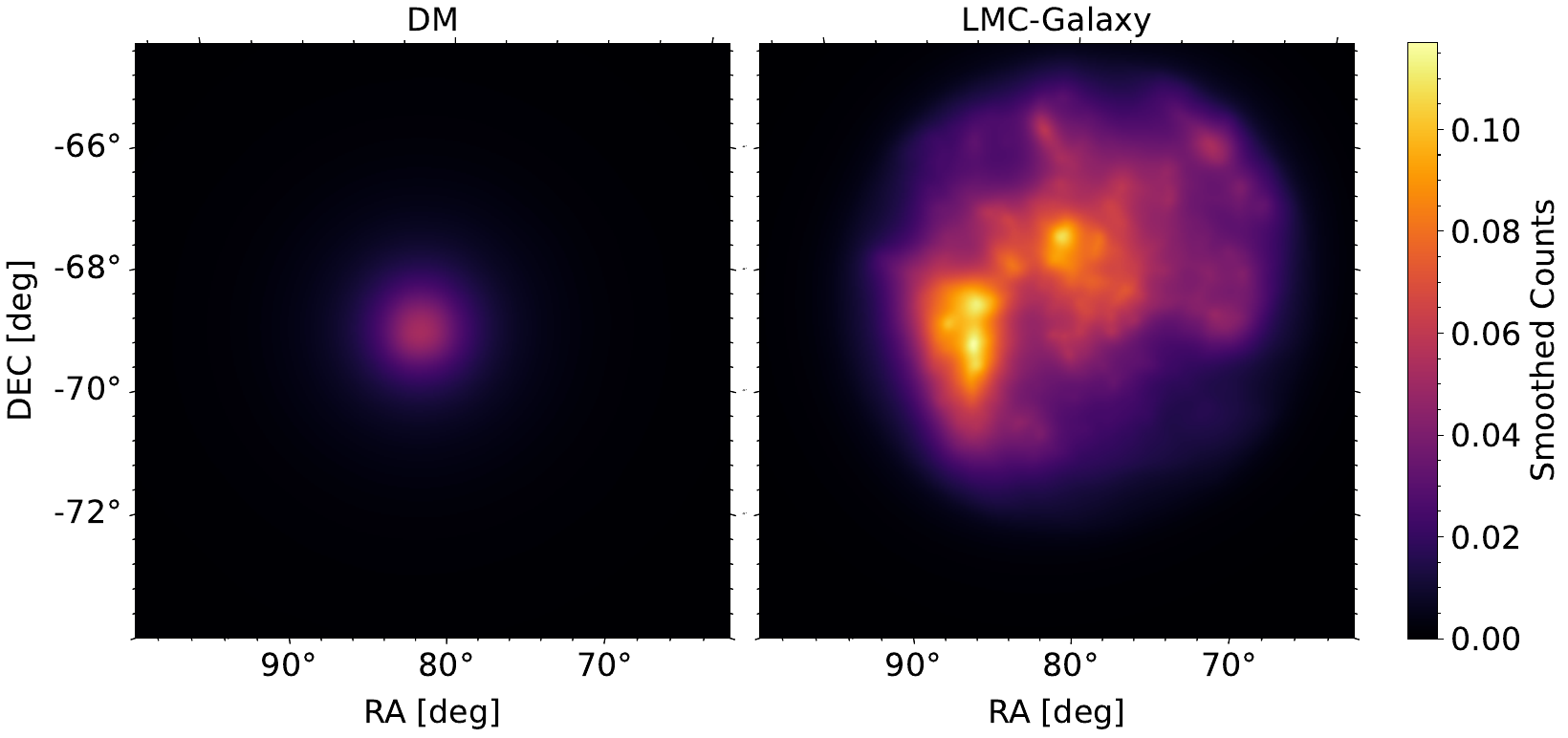}
    \caption{Left: Model counts maps of our added LMC DM source assuming s-wave annihilation. The counts of our added DM source assumes a DM mass of 30 GeV, annihilation cross section of $\sigma v_{\rm rel} = 7 \times 10^{-27}$ cm$^{3}$ s$^{-1}$, and annihilation into $b \bar b$ channel. Right: The LMC-Galaxy source ($\texttt{4FGL J0519.9-6845e}$). Figures are both smoothed by the instrumental PSF. \label{lmc_counts}}
\end{figure}

In Figure \ref{lmc_counts} we compare the model counts map of our added LMC DM source with that of the LMC-Galaxy source. The counts of the DM source are weak when compared with the counts of the LMC-Galaxy source. The $\mathcal{J}$-factor maps from the simulation were used as the spatial model for our added DM source.

Given the shape of our LMC simulation template, we have one free model parameter, which is the normalization of the emission from this component. This emission is proportional to the DM annihilation cross section, with the shape of the spectrum determined by the assumed DM mass and annihilation channel. For our simple exercise, we assume a DM mass of 30 GeV, and a $b\bar{b}$ channel annihilation spectrum. For these assumptions, we find a TS of $\sim 138$ for the LMC DM source.

Although this may be indicative of a detected source, it is more likely a reflection of the complicated emission within the region. For example, if we were to remove the LMC DM source from the map, the total likelihood for the fit improves without the DM source.

\begin{figure}[t]
    \centering
    \includegraphics[width=0.8\linewidth,clip]{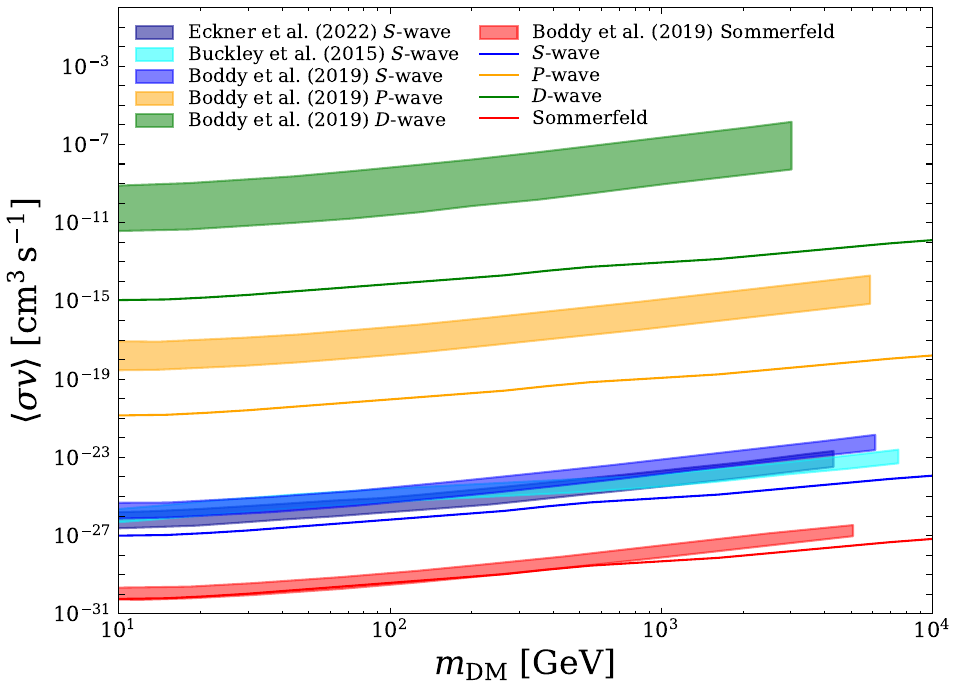}
    \caption{Gamma-ray upper limits on the annihilation cross section of DM particles in the LMC for s-wave, p-wave, d-wave, and Sommerfeld, assuming $b \bar b$ annihilation. We calculate the s-wave limits, then scale them by the relative ratio of the ${\cal J}$-factor for each model and the s-wave ${\cal J}$-factor from the simulation maps to obtain the p-wave, d-wave, and Sommerfeld limits. Also shown are comparisons to previous LMC limits derived from Fermi-LAT data~\cite{Eckner:2022swf, Buckley2015, Boddy:2019qak}.\label{lmc_exclusion}}
\end{figure}

Given the ambiguity of the source detection, we can consider the sensitivity to the LMC signal in the form of cross section upper limits, which are shown in Figure~\ref{lmc_exclusion}. For  s-wave, the bounds we derive are consistent with those from previous studies \cite{Eckner:2022swf}, \cite{Buckley2015}, and \cite{Boddy:2019qak}, given the differences in the ${\cal J}$-factors used for the LMC. Our simulation template allows us to place the first bounds on velocity-dependent annihilation cross sections from the LMC. To compute the bounds for the p-wave, d-wave, and Sommerfeld models, we scale the s-wave limit by the ratio of the ${\cal J}$-factor from each model and the s-wave  ${\cal J}$-factor. It is interesting to note that our bounds from the LMC are  $\sim 4-6$ orders of magnitude more stringent than previous dwarf galaxy bounds for p-wave emission. For d-wave emission, the bounds are $\sim 2-3$ orders of magnitude more stringent than previous studies.

\section{Discussion and conclusions}
\label{sec: discussion}

In this work we have studied the DM annihilation signals from the LMC and the impact of the LMC on DM annihilation signals from the MW halo, using a simulated MW-LMC analogue from the Auriga magneto-hydrodynamical simulations. We have considered the velocity-independent s-wave DM annihilation, as well as the Sommerfeld, p-wave, and d-wave models in which the annihilation cross section is velocity-dependent.

Our computation of the LMC $\mathcal{J}$-factor demonstrated that the emission from the LMC rises above the MW foreground by a factor of more than 100 for the four annihilation models  considered. We found that the angular size of the region around the LMC at which the DM annihilation signal is observable above the MW foreground is largest for Sommerfeld-enhanced annihilation and smallest for d-wave annihilation. 

We have also quantified the impact of the LMC on the $\mathcal{J}$-factor from the MW halo by comparing two simulation snapshots: the isolated MW analogue (before the infall of the LMC into the MW), and the present day MW-LMC analogue. We found that the impact of the LMC on the MW $\mathcal{J}$-factor is minimal for Sommerfeld and s-wave annihilation. For the case of p-wave and d-wave annihilation, the $\mathcal{J}$-factor is  boosted by up to a factor of $\sim 4$ and $\sim 6$, respectively. In particular, high-speed DM particles brought into the MW halo with the LMC as well as the boosted MW DM particles due
to the passage of the LMC impact the $\mathcal{J}$-factors for these annihilation models. Additionally, the LMC has the largest effect on the MW in the outer halo where the increase in the DM density and relative velocity distribution is the largest.

Using over 15 years of Fermi-LAT data, we analyze bounds on the DM annihilation cross section from the LMC. While the emission from the DM component of the LMC is smooth and extended, it is still less extended and morphologically less complex than the astrophysical emission from the LMC. We show that it is possible to find a fit that includes emission from a DM component; however the likelihood for this fit implies that it is disfavored relative to the fit that only includes the astrophysical emission from the LMC. We obtain bounds on the cross section consistent with, and in some cases more stringent than those presented in refs.~\cite{Eckner:2022swf, Buckley2015, Boddy:2019qak}, accounting for the differences in our ${\cal J}$-factors and  the mass of the LMC analogue. Bounds for d-wave annihilation are  $\sim 4-6$ orders of magnitude more stringent relative to published bounds from dwarf galaxies, and $\sim 2-3$ orders of magnitude more stringent for p-wave models.

Our results demonstrate that the impact of the LMC on the DM velocity distribution in the MW halo is non-negligible, and this affects the DM annihilation signals for the p-wave and d-wave velocity-dependent annihilation models in a non-trivial way. Future high resolution cosmological simulations with more MW-LMC analogues would help further quantify the impact of the LMC on the MW halo.


\acknowledgments

 We thank Azadeh Fattahi for providing the re-simulated Auriga halo studied in this work. EV and NB acknowledge the support of the Natural Sciences and Engineering Research Council of Canada (NSERC), funding reference number RGPIN-2020-07138 and the NSERC Discovery Launch Supplement, DGECR-2020-00231. AE, OH, and LS acknowledge support from DOE Grant de-sc0010813, NASA grant 80NSSC22K1577, and by the Texas A\&M University System National Laboratories Office and Los Alamos National Laboratory. NB acknowledges the support of the Canada Research Chairs Program. 
 


\appendix

\section{Variations in the MW $\mathcal{J}$-factor}
\label{app: var}

In this appendix we discuss the variations in the MW $\mathcal{J}$-factor boost due to the LMC when changing the Solar position.

In section~\ref{sec: MWJ}, the Solar position was chosen in order to best match the on-sky position of the LMC analogue to its observed position (as discussed in section~\ref{sec: Solar}). This is particularly important when computing the $\mathcal{J}$-factor for the LMC itself, in order to ensure that the MW foreground in the direction of the LMC analogue is accurate. This Solar position also results in the LMC analogue being at a distance of $\sim 51$~kpc and it's orbit passing below the MW center, both of which are in agreement with observations. However, it is difficult to determine how the DM distribution around the Solar position depends on the correct LMC on-sky position, distance and orbit. Varying the Solar position from its best matched position may lead to a different local distribution of LMC DM particles as well as boosted MW DM particles. Nevertheless, to understand the variations in the MW $\mathcal{J}$-factor boost, it is interesting to investigate how the local DM distribution and hence the $\mathcal{J}$-factor boost in the outer halo varies for different Solar positions. 

To check this, we consider three rings aligned with the stellar disk and centered on the MW center with radii of 7, 8 and 9 kpc. We then choose 10 equally spaced Solar positions around each ring, resulting in a total of 30 Solar positions for the three rings. For each Solar position, we compute the average $\mathcal{J}$-factor boost in the direction of the MW anticenter in a portion of the sky that is $30^\circ \times 30^\circ$. This provides a range of $\mathcal{J}$-factor boosts that one can expect from the presence of the LMC, depending on the position of the Sun in the stellar disk.

We compute the average boost for each of the 30 Solar positions.  The mean value of these average boosts along with its standard error are $0.96 \pm 0.05$, $1.19 \pm 0.05$, $1.76 \pm 0.08$ and $2.55 \pm 0.13$ for the Sommerfeld, s-wave, p-wave and d-wave models, respectively. The minimum/maximum of the mean values are 0.66/1.79, 0.84/1.84, 1.11/2.65 and 1.48/4.03 for the same four models, respectively. For the Sommerfeld model, we find that the presence of the LMC generally reduces the $\mathcal{J}$-factor in the MW outer halo. This is because the LMC introduces high-speed DM particles into the MW and also boosts the native MW DM particles, shifting the relative DM velocity distribution in the MW halo to higher speeds. For Sommerfeld models, the annihilation of fast-moving DM particles is suppressed relative to slower-moving DM. For the case of s-wave, we find that the outer MW $\mathcal{J}$-factor is slightly boosted. This is a result of the DM particles that fall into the MW with the LMC as well as the change in the DM density in the MW halo due to the passage of the LMC. For p-wave and d-wave models, the outer MW $\mathcal{J}$-factor is boosted more so than for s-wave since these models are sensitive to the high speed tail of the DM relative velocity distribution and the  annihilation of fast-moving DM particles is enhanced for them. 

Finally, notice that the maximum/minimum values of the mean boosts for Sommerfeld and s-wave show that depending on the Solar position, the outer MW halo $\mathcal{J}$-factor may be larger or smaller in the $\textit{Pres.}$ snapshot compared to the $\textit{Iso.}$ snapshot. This is in contrast to p-wave and d-wave, where there is always a boost. For Sommerfeld, despite the average $\mathcal{J}$-factor ratio being lower than 1, the maximum $\mathcal{J}$-factor ratio is larger than 1. This indicates that for specific Solar positions, the increase in the number of DM particles in the direction of the MW anticenter due to the LMC dominates over the suppression of fast-moving DM particles. For s-wave, despite the average $\mathcal{J}$-factor ratio being larger than 1, the minimum $\mathcal{J}$-factor ratio is smaller than 1. This suggests that for specific Solar positions, the line-of-sight DM density in the direction of the MW anticenter is larger in the $\textit{Iso.}$ snapshot compared to the $\textit{Pres.}$ snapshot. This may  be due to natural variations between the halo in the two snapshots for this particular line-of-sight. On the other hand, it may indicate an LMC-induced underdensity in the DM distribution for those specific Solar positions. 

\clearpage

\bibliographystyle{JHEP}
\bibliography{refs}

\end{document}